\documentclass{article}

\usepackage{arxiv}

\usepackage[utf8]{inputenc}
\usepackage[T1]{fontenc}
\usepackage{hyperref}
\usepackage{url}
\usepackage{booktabs}
\usepackage{amsfonts}
\usepackage{amsmath}
\usepackage{amssymb}
\usepackage{nicefrac}
\usepackage{microtype}
\usepackage{cleveref}
\usepackage{graphicx}
\usepackage{natbib}
\usepackage{doi}
\usepackage{array}
\usepackage{booktabs}
\usepackage{multirow}
\usepackage{tipa}

\title{Music-Source-Separation-Training (MSST): A Unified Framework for Training and Evaluating Music Demixing Models}
\date{July 17, 2026}

\newif\ifuniqueAffiliation
\uniqueAffiliationtrue

\ifuniqueAffiliation
\author{Roman Solovyev \\
	MVSep.com \\
	\texttt{zfturbo@gmail.com} \\
	\And
	Ilya Kiselev \\
	National Research University Higher School of Economics (HSE University), MIEM HSE \\
	\texttt{ilkiselya@gmail.com} \\
	\And
	Alexander Stempkovskiy \\
	AlphaChip LLC.\\
	\texttt{stemp@ippm.ru}
	\And
	Tatiana Gabruseva \\
	Independent Researcher \\
	\texttt{tatigabru@gmail.com} \\
}
\fi

\renewcommand{\shorttitle}{Music-Source-Separation-Training}

\hypersetup{
pdftitle={Music-Source-Separation-Training: A Unified Framework for Training and Evaluating Music Demixing Models},
pdfsubject={cs.SD, eess.AS},
pdfauthor={Roman Solovyev, Tatiana Gabruseva, Ilya Kiselev, Alexander Stempkovskiy},
pdfkeywords={music source separation, deep learning, audio processing},
}

\begin{document}
\maketitle

\begin{abstract}
Music Source Separation (MSS), the task of recovering individual sound components (stems) from
a polyphonic mixture, is central to applications ranging from karaoke and remixing to audio
restoration and content production. The separation quality depends on engineering decisions
across the entire pipeline: model choice, training data preparation and augmentation,
loss function and metrics choice, training configuration, validation, and post-processing.

This paper presents MSST (Music-Source-Separation-Training) -- a universal open-source framework for MSS tasks,
which unifies training, validation, and inference
for a broad range of modern demixing model families under a single, configuration-driven interface.
The framework supports various model architectures, data preprocessing and augmentations, multiple
loss functions and evaluation metrics, which helps with fast iterations and ablation studies. Additionally,
the framework supports a range of practical techniques that improve separation quality, such as sliding-window inference
with cross-fading, test-time augmentation, model ensembling, and fine-tuning via Low-Rank Adaptation (LoRA).
Our ablation studies demonstrate improvements of MSS using the above techniques. By consolidating these
components into a reproducible, YAML-configurable framework, MSST lowers the barrier to systematic
experimentation and enables rapid iteration from idea to verifiable result.
\end{abstract}


\section{Introduction}

Music Source Separation (MSS) is a core problem in audio machine learning. 
Its goal is to decompose a polyphonic mix into its constituent components -- 
vocals, drums, bass, and other instruments (commonly called ``stems''). 
MSS underpins a wide range of applications, including karaoke and remixing tools, 
audio restoration and remastering, music information retrieval, content production, and audio engineering workflows.

MSS poses substantial engineering challenges. Models must handle heavily overlapping signals in both frequency and time, preserve phase coherence, remain robust to artefacts, and tolerate the variability introduced by different mixing and mastering practices. 
Modern MSS solutions rely primarily on deep neural networks and exploit different signal representations: the time domain (raw waveforms) \cite{defossez2019demucs, waveunet}, the spectral domain (STFT spectrograms) \cite{stoter2019umx}, and hybrid combinations of both \cite{defossez2021hybrid, rouard2023htdemucs}.

In production systems, however, model architecture choice is only one of the factors among many; the full pipeline determines the separation output quality, including data preparation, augmentation strategies, loss function and evaluation metrics selection, inference strategies for long audio files, and ease of integration into downstream workflows. Finding the optimal pipeline choices usually requires rapid iterations and experimentation with different parameters of the entire pipeline.

In this paper we introduce Music-Source-Separation-Training (MSST) \cite{msst_github}, 
an open-source framework that unifies training, validation, and inference for a wide range 
of modern model families, including Band-Split RoFormer, Mel-Band RoFormer, HTDemucs, SCNet, 
and others. We describe protocols and engineering decisions that directly affect separation quality and usability: data augmentations, different loss functions and their combination, evaluation metrics, training across multiple GPUs, test-time augmentation (TTA), model ensembling, and LoRA fine-tuning. Our ablation studies demonstrate reliable improvements of MSS using the proposed techniques.

\section{Related Work}

MSS training codebases fall into broad categories: (a) single-model research repositories (e.g., Demucs), 
(b) reference implementations with end-to-end pipelines for a narrow task (e.g., Open-Unmix), and 
(c) general-purpose source separation toolkits (e.g., Asteroid).

Here, we give an overview of the relevant open-sourced libraries and for MSS training (as of July 2026):

\begin{itemize}
	\item \textbf{Open-Unmix} \cite{stoter2019umx, sigsep_openunmix_github}: A minimal 
	PyTorch reference implementation for MSS, aimed at researchers, audio engineers, and artists. 
	Each target is separated by an independent three-layer bidirectional LSTM operating on magnitude 
	spectrograms, with an optional expectation-maximisation / multichannel Wiener-filter post-processing 
	stage. It ships ready-to-use models for four-stems separation (vocals, drums, bass, and the other) pre-trained on the MUSDB18 
	and MUSDB18-HQ datasets, a \texttt{umxl} trained with extra data (non-commercial licence), 
	and a Sony-trained speech-enhancement model (\texttt{umxse}). 
	Distributed as a pip package with weights on Zenodo and Torch Hub, 
	it prioritises readability and reproducibility over breadth of architectures, 
	and remains lightly maintained (updated for Torch~2.0 in 2024). 
	It served as an official baseline for the 2021 MDX Challenge \cite{MDX21}. 

	\item \textbf{Spleeter} \cite{Spleeter, deezer_spleeter_github}: 
	A TensorFlow-based MSS separator built around encoder/decoder U-Net models with skip connections, 
	one per target, that estimate soft (or multichannel Wiener) masks over the spectrogram. 
	It ships 2-stem (vocals vs accompaniment), 4-stem (vocals, drums, bass, and the other), 
	and 5-stem (adding piano) pre-trained models and is engineered for the fast inference throughput. 
	The pre-processing and estimator are implemented with the TensorFlow 
	\texttt{data} and \texttt{estimator} APIs, with spectrogram caching in train/eval and a fused 
	separation graph at predict time. Training is supported via a JSON model configuration 
	and a \texttt{spleeter train} CLI, though the training \emph{data} is not released. 
	It is MIT-licensed, distributed via pip/conda/Docker, and heavily used in downstream audio software.

	\item \textbf{Demucs} (Facebook AI Research) 
	\cite{defossez2019demucs, rouard2023htdemucs, facebookresearch_demucs_github}: 
	A state-of-the-art architecture family originating from a waveform-domain U-Net inspired 
	by Wave-U-Net. The v4 release introduces Hybrid Transformer Demucs (HTDemucs), which adds a 
	spectrogram branch (using complex-as-channels or masked magnitude) alongside the temporal branch 
	and inserts cross-domain Transformer blocks; a Sparse HT variant and an experimental six-source 
	model (adding guitar and piano) are also described. 
	It provides multiple pre-trained models (\texttt{htdemucs}, \texttt{htdemucs\_ft}, \texttt{hdemucs\_mmi}, \texttt{mdx}, \texttt{mdx\_extra}) 
	and full training code driven by the Dora experiment manager. 
	Notably, the upstream repository was archived (read-only) in January~2025, 
	with continued work moved to a personal fork.

	\item \textbf{Asteroid} \cite{pariente2020asteroid, asteroid_github}: 
	The most prominent general-purpose PyTorch source separation toolkit (not limited to music), 
	built on PyTorch and PyTorch-Lightning. It exposes composable building blocks, such as filterbanks, 
	encoders, maskers, decoders, and losses, that can be recombined into new systems, 
	and ships bash recipes covering data preparation, training, and evaluation across many speech- 
	and music-separation datasets. Pretrained models are hosted on Zenodo with a sharing CLI. 
	Its orientation is academic and reproducibility-focused, spanning speech enhancement, 
	speech separation, and music, rather than tracking the latest music-specific SOTA architectures.

	\item \textbf{KUIELab-MDX-Net} \cite{Kim21, kuielab_mdxnet_github}: 
	The official training repository for a top-placing MDX~2021 system 
	(2nd on Leaderboard~A, 3rd on Leaderboard~B). It has a two-stream design: 
	a spectrogram branch built from per-source, enhanced TFC-TDF U-Nets, blended with 
	a waveform branch (the original Demucs), followed by a lightweight 1$\times$1-convolution 
	``Mixer'' that refines the four estimated stems jointly. 
	Built on a PyTorch Lightning-Hydra template with Hydra/YAML experiment configs, 
	it provides per-source training scripts and offline data augmentation 
	(pitch/tempo shift, random track mixing). It is explicitly referenced by the MDX Challenge 
	paper as providing training code.

	\item \textbf{nussl} (Northwestern University) \cite{manilow2018nussl, nussl_github}: 
	A historically significant, object-oriented PyTorch separation library from the Interactive 
	Audio Lab, spanning classical primitives (REPET, 2DFT, harmonic/percussive, DUET, PROJET) and 
	deep-learning models with a built-in trainer. It offers dataset hooks (MUSDB18, WSJ0, WHAM!, FUSS), 
	Scaper-based augmentation, evaluation (BSSEvalv4/\texttt{museval}, SI-SDR and variants), 
	and a model zoo (External File Zoo). 
	The repository is now archived (read-only since December~2024).

	\item \textbf{Danna-Sep} \cite{yu2021dannasep, dannasep_github}: 
	An open repository for a high-ranking MDX~2021 system that linearly blends three 
	complementary sub-models across feature domains: a modified X-UMX and a U-Net 
	(both spectrogram-based) plus an enhanced Demucs (waveform-based), exploiting the 
	observation that spectrogram models favour harmonic sources while waveform models 
	favour percussive/bass content. Separate inference and training repositories are provided, 
	with a pip-installable CLI that downloads $\sim$1--2\,GB of pre-trained weights and outputs 
	four stems.

	\item \textbf{DeepConvSep} \cite{chandna2017deepconvsep, deepconvsep_github}: 
	An older but explicitly training-oriented music-separation codebase implemented 
	in Lasagne/Theano. It provides STFT feature computation and convolutional encoder/decoder 
	models that predict time-frequency soft masks, with dataset-specific training 
	pipelines for iKala (singing-voice separation), DSD100 (voice/bass/drums/accompaniment), 
	and Bach10 (wind/string instruments, optionally score-informed via the RWC instrument database). 
	Evaluation is provided as MATLAB BSS-Eval scripts. It predates the PyTorch/TensorFlow 
	ecosystem and is of primarily historical interest.

	\item \textbf{Bytedance music\_source\_separation} (ByteDance) 
	\cite{kong2021bytedance, bytedance_mss_github}: 
	A PyTorch codebase (\texttt{bytesep}) that supports training custom separation systems in 
	addition to inference, based on a decoupled magnitude/phase ResUNet from the associated 
	ISMIR~2021 work \cite{kong2021bytedance}. 
	It exposes fine-tuning and resume-from-checkpoint templates and is applicable 
	beyond music (e.g. speech enhancement, instrument separation), 
	though its benchmark positioning and current maintenance status are not fully documented.

	\end{itemize}

The comparison below emphasises the most established toolkits and 
benchmark-anchored codebases while noting influential archived projects.
Table~\ref{tab:comparison} summarises the key features of the surveyed libraries. 
Among these, the Demucs and nussl repositories are no longer actively maintained: 
the upstream Demucs repository was archived (read-only) on 1~January~2025, and nussl has 
been archived since December~2024; the remaining libraries are, to varying degrees, still maintained.

\begin{table*}[ht]
\caption{Comparison of open-source MSS training libraries.}
\label{tab:comparison}
\resizebox{\textwidth}{!}{%
\begin{tabular}{p{5.0cm}p{2.2cm}p{6.8cm}p{1.6cm}}
\toprule
\textbf{Library / Scope} & \textbf{Framework} & \textbf{Features} & \textbf{Licence} \\
\midrule
MSST; MSS training \& inference & Python/PyTorch & Training and inference support, multiple model families, configurable losses, augmentations, metrics, datasets, pretrained support, DDP/AMP. & MIT \\
Open-Unmix; 4-stem demixing & Python/PyTorch & Reference 4-stem pipeline, basic training support, pretrained models, benchmark dataset support. & \texttt{--} \\
Spleeter; Fast separator \& training & Python/TensorFlow & Fast separator and training, 4-stem models, simple dataset handling. & \texttt{--} \\
Demucs; SOTA model family & Python/PyTorch & Demucs model famyily (hybrid waveform models), pretrained checkpoints, training recipes, benchmark. & \texttt{--} \\
Asteroid; General toolkit \& recipes & Python/PyTorch & Modular toolkit, training recipes, losses, metrics, dataset wrappers. & \texttt{--} \\
KUIELab-MDX-Net & Python/PyTorch & MDX challenge models, training pipeline, dataset support. & \texttt{--} \\
nussl; General separation research & Python & Research library for general object-oriented separation, training and evaluation hooks. & MIT \\
Danna-Sep & Python/PyTorch & Competition-oriented hybrid model, training workflow, dataset-specific support. & MIT \\
DeepConvSep; CNN training code & Python & CNN models, training scripts, dataset-specific workflows. & \texttt{--} \\
Bytedance MSS; MSS training \& inference & Python/PyTorch & Training and inference support, model family, dataset handling. & \texttt{--} \\
\bottomrule
\end{tabular}%
}
\end{table*}

MSST is a universal training and inference framework for MSS tasks, which is designed for fast experiment 
velocity. It supports a wide choice of modern model architectures, easy model swapping via YAML-based 
configurations, different dataset types and configurable data processing and augmentations, 
flexible loss function composition, multiple validation metrics, together with practical engineering features such 
as distributed data parallel training, mixed precision, exponential moving average, weighted ensembling, 
optional LoRA fine-tuning, and test-time augmentation. Whereas many popular open-source MSS projects focus on a single architecture family or a specific separation 
setting, MSST supports a wide range of model families behind a single training and validation loop, 
whch is easily configurable for high-throughput ablation studies.

\section{MSS Problem Formulation and Approaches}

Here, we discuss mathematical formulations for the MSS task, which involves recovering individual sound components (stems) from a mixture.
Mathematically, a mixed audio signal (the mix) is represented as an additive combination of $N$ source signals (stems):

\begin{equation}
    x(t) = \sum_{i=1}^{N} s_i(t)
\end{equation}

where $s_i(t)$ denotes the signal of source $i$ and $x(t)$ is the observed mix.

The goal of MSS is to recover accurate estimates of the individual components $\hat{s}_i(t)$ from $x(t)$. Ideally, the sum of the estimated sources should reconstruct the original mix without distortion. Models are typically trained in a supervised fashion using datasets of (mix, ground-truth stems) pairs obtained from multitrack studio sessions. A notable limitation of this setup is that real-world mixes undergo a mastering stage involving nonlinear processing, whereas model training typically assumes a simplified linear summation of individual tracks.

Three principal approaches to MSS are recognised in current practice:

\begin{itemize}
	\item \textbf{Spectrogram masking.} The model analyses the mix spectrogram obtained via the Short-Time Fourier Transform (STFT) and predicts a mask for each target source --- a weight matrix with values in $[0, 1]$ applied element-wise to the mix spectrogram. For example, Open-Unmix \cite{stoter2019umx} uses a bidirectional LSTM to predict spectral masks for vocals and accompaniment. This approach reuses the mix phase unchanged and focuses on estimating spectral magnitudes for each source. It is computationally efficient but limited in achievable quality because a single phase is shared across all recovered sources.

	\item \textbf{Direct signal regression.} Models operate directly in the time domain, predicting the waveform of each stem from the input mix. Wave-U-Net \cite{waveunet} exemplifies this class: the model performs complex nonlinear filtering of the audio signal using convolutional or recurrent layers. Direct waveform prediction avoids the need for separate phase handling, as the signal is reconstructed in full. However, such models are more difficult to train because they must reproduce signal oscillations at high sampling rates.

	\item \textbf{Hybrid methods.} The most promising direction combines the strengths of both approaches. These models operate in the frequency domain but are optimised with respect to temporal and phase characteristics. In MSST, \textbf{Band-Split RoFormer} (\texttt{bs\_roformer}) \cite{roformer} and \textbf{Mel-Band RoFormer} (\texttt{mel\_band\_roformer}) \cite{wang2023mel} belong to this class. These models process spectral representations by splitting the frequency axis into bands (linear or mel-scale) and applying Transformer blocks to each band independently. The output is not a binary mask; instead, the model predicts flexible spectral transformations, and signal reconstruction incorporates phase information and spectral losses. \textbf{SCNet} \cite{scnet} also belongs to this class: it operates on the complex spectrogram and predicts values that approximate direct regression of the source spectrum rather than a pure mask. The hybrid approach better captures both local temporal details (attacks, transients) and global spectral structure, yielding more consistent separation quality in practice.
\end{itemize}

\begin{table}[ht]
\caption{Comparison of MSS approaches.}
\label{tab:approaches}
\centering
\small
\begin{tabular}{p{2.4cm}p{3.2cm}p{3.2cm}p{3.2cm}}
\toprule
\textbf{Criterion} & \textbf{Spectrogram Masking} & \textbf{Direct Regression} & \textbf{Hybrid Methods} \\
\midrule
Processing domain & Frequency (STFT) & Time (raw audio) & Time-frequency / Complex \\
Phase handling & Reuses mix phase & Reconstructs phase directly & Optimises phase or uses complex weights \\
Advantages & High speed, stability, good timbre separation & No phase artefacts, better transient reproduction & State-of-the-art quality, minimal artefacts, versatility \\
Disadvantages & Quality ceiling due to shared mix phase & Difficult to train, high resource requirements & High architectural complexity, slow inference \\
Representative models & Open-Unmix, Spleeter & Wave-U-Net, Conv-TasNet & BS-RoFormer, SCNet, HTDemucs \\
\bottomrule
\end{tabular}
\end{table}

\section{Model Architectures}

Modern MSS models employ diverse architectural patterns, from convolutional 
networks to Transformers. The MSST framework supports a broad set of model families, including the following:

\begin{itemize}
	\item \textbf{TFC-TDF U-Net (MDX23C)} \cite{MDX23C}. A two-stream U-Net architecture proposed for the Sound Demixing Challenge 2023. The network simultaneously extracts features in the time domain and on the spectrogram (TDF --- Time-Distributed Fourier, TFC --- Time-Frequency Convolution) and splits the spectrum into bands, improving the separation of low and high frequencies.

	\item \textbf{Demucs / HTDemucs} \cite{rouard2023htdemucs}. A family of U-Net models operating directly on waveform signals. The hybrid version, Hybrid Transformer Demucs (HTDemucs), combines a convolutional encoder-decoder with Transformer blocks, improving high-frequency reconstruction and increasing SDR when additional training data is available.

	\item \textbf{Band-Split RoFormer} \cite{roformer}. A Transformer architecture that partitions the spectrum into frequency bands, applies a RoFormer with Rotary Position Embeddings to each band, and merges the results. This model achieved state-of-the-art results on the SDX23 benchmark.

	\item \textbf{Mel-Band RoFormer} \cite{wang2023mel}. An extension of BS-RoFormer in which the bands are defined on the mel scale, improving separation quality for vocals, drums, and other instruments.

	\item \textbf{SCNet and variants} \cite{scnet}. A complex-valued U-Net operating on the real and imaginary parts of the spectrogram. The \texttt{scnet\_tran} and \texttt{scnet\_masked} variants add Transformer blocks and masking for different output modes.

	\item \textbf{Bandit / Bandit-Plus} \cite{watcharasupat2023bandit}. A Bandsplit U-Net that generalises Bandsplit-RNN using a shared encoder, psychoacoustic frequency scales, and a novel SNR-motivated loss function to improve performance and reduce computational complexity.

	\item \textbf{Bandit v2} \cite{mvsep2023banditv2}. A model designed for three-stem separation (speech, music, effects), trained on the Divide \& Remaster v3 dataset.

	\item \textbf{BSMamba2} \cite{kim2024bsmamba2}. A Mamba2-based model with a band-splitting and dual-path scheme capable of processing long sequences, demonstrating high SDR and robustness to varying input lengths.

	\item \textbf{Conformer / BS Conformer} \cite{gulati2020conformer}. Conformer combines convolutional and self-attention blocks to capture both local and global dependencies. The BS Conformer variant transfers this design to MSS by splitting the spectrum into bands, analogous to BS-RoFormer.

	\item \textbf{Swin Upernet} \cite{liu2021swin}. A U-Net built on the Swin Transformer (a Vision Transformer with shifted windows) and a UperNet decoder head, providing hierarchical spectrogram modelling as a backbone for MSS.

	\item \textbf{Segmentation-based models (VitLarge23, TorchSeg).} Experimental models that use pretrained Vision Transformer (ViT) backbones and the TorchSeg library for spectrogram segmentation, demonstrating the transfer of computer vision methods to MSS.

	\item \textbf{Other experimental models.} The repository contains variants such as SCNet-Tran, SCNet-Masked, and Band-Split RoFormer with additional parameters and alternative backbone combinations. These enable investigation of the effects of self-attention, sub-band partitioning, and masking on separation quality.
\end{itemize}

\section{Datasets and Data Preparation}

\subsection{Data Collection and the Concept of a Stem}

Training MSS models requires multitrack audio sessions containing individual tracks for vocals, drums, bass, and other instruments, which together form the final mix. Each isolated track is called a \textbf{stem}. Obtaining stems from commercial recordings is constrained by limited access to source material and by licensing restrictions. Nevertheless, the community has produced several open datasets for MSS research:

\begin{itemize}
	\item \textbf{MUSDB18} \cite{musdb18}. The standard benchmark dataset consisting of 150 tracks ($\approx$10 hours) across various genres, split into four stems: vocals, drums, bass, and other instruments. The data is divided into training (100 songs) and test (50 songs) subsets. Audio is provided as stereo WAV/MP4 at 44.1~kHz. MUSDB18 has become the primary evaluation benchmark for MSS algorithms.

	\item \textbf{MoisesDB} \cite{moises}. A dataset released by Moises in 2023 containing 240 professionally recorded songs (45 artists, 12 genres) with finer-grained source separation (up to 5--6, and in some cases 11 stems per track). MoisesDB enables training models to separate guitars, keyboards, and other instruments individually, extending beyond the standard four-stem configuration.

	\item \textbf{Other and custom datasets.} Demixing tasks also make use of multitrack sessions from adjacent domains, such as audio tracks from video games or studio STEM releases. MSST allows multiple audio folders to be specified for training (the \texttt{--data\_path} argument accepts multiple paths), enabling the combination of heterogeneous data sources. Custom datasets require no additional programming --- it is sufficient to organise files in one of the supported directory structures and specify paths in the configuration.
\end{itemize}

\subsection{Dataset Types}

MSST supports seven dataset types, each suited to different data organisation needs:

\begin{itemize}
	\item \textbf{Type 1 (MUSDB format).} Each song resides in a separate folder containing stem files named \texttt{<stem\_name>.wav} (or \texttt{.flac}). This is the base format of MUSDB18/MUSDB18-HQ --- the simplest and most universal format for mix-to-stems separation.

	\item \textbf{Type 2 (Stems folders).} Each folder corresponds to a single instrument class (\texttt{vocals/}, \texttt{drums/}, \ldots) and contains clean recordings of that stem only. Mixes are assembled on the fly during training.

	\item \textbf{Type 3 (CSV indexing).} Instead of a fixed directory structure, a CSV file with columns \texttt{instrum, path} assigns a class label and file path to each track. This enables flexible combination of data from different directories and datasets without moving files.

	\item \textbf{Type 4 (MUSDB Aligned).} The same structure as Type 1, but during training all stems are loaded from the same temporal position within a track (aligned chunks). This ensures correct stem synchronisation when training on chunks, which is important for models and losses sensitive to temporal alignment.

	\item \textbf{Type 5 (Precomputed Chunks).} The same structure as Type 4, but tracks are pre-split into chunks with 50\% overlap and stored on disk. This provides faster and more stable I/O, useful when on-the-fly chunking is computationally expensive or when a fully deterministic dataset is desired.

	\item \textbf{Type 6 (Aligned + Explicit Mixture).} An extension of Type 4 in which each folder may contain a \texttt{mixture.wav} file. If present, the mix is loaded directly; if absent, it is reconstructed as the sum of the stems. This is important for scenarios where the real mix is not equal to the sum of stems (due to mastering effects), and for teacher--student, distillation, and consistency-loss approaches.

	\item \textbf{Type 7 (Class-Balanced Aligned).} The same structure as Type 6, but with a sampling strategy designed to reduce class frequency imbalance and improve learning of rare instruments. The loader first samples a class, then selects a track containing that class, and loads an aligned chunk; it additionally returns a list of active stems (\texttt{active\_stem\_ids}). This is beneficial for sparse multi-instrument datasets and tasks involving rare instruments or conditional separation.
\end{itemize}

\subsection{Data Augmentations}

To make effective use of limited data volumes, MSST applies data augmentation extensively, modifying audio tracks on the fly to generate new mix variants. Augmentations can be configured globally for the entire mix or independently for each stem. The supported techniques include:

\begin{itemize}
	\item \textbf{Stereo and temporal transformations}: channel shuffling, time reversal, and polarity inversion.

	\item \textbf{Pitch and tempo modifications}: pitch shifting (in semitones) and time stretching (speed variation with preserved segment length), via \texttt{audiomentations}~\cite{audiomentations}.

	\item \textbf{Processing and codec artefact simulation}: a parametric seven-band equaliser, tanh distortion, Gaussian noise injection, and MP3 compression.

	\item \textbf{Pedalboard effects}: reverberation, chorus, phaser, distortion, bitcrush, resampling, MP3 compressor, and an alternative pitch shift implementation, via \texttt{pedalboard}~\cite{pedalboard}.
\end{itemize}

Augmentations are configured via the YAML config file (the \texttt{augmentations} section). Separate transformation chains can be specified for different sources --- for example, applying pitch shifting only to instrumental tracks and reverberation only to vocals. This flexibility allows the training distribution to approximate real-world conditions more closely and helps prevent overfitting on small datasets.

\subsection{Data Preprocessing}

The repository provides utilities for dataset preparation. The script \texttt{utils/dataset.py} reads specified audio folders, splits tracks into training segments of a given length, and mixes corresponding stems into a summed signal. It also supports combining tracks from different songs to generate synthetic mixes. A validation set path can be specified via the \texttt{--valid\_path} parameter. The data loading process is fully configuration-driven, and custom datasets can be integrated without code changes.

\section{Training}

\subsection{Loss Functions}

The file \texttt{utils/losses.py} implements a set of loss functions that can be combined as a weighted sum (weights are set via command-line parameters and the config). MSST supports time-domain losses as well as spectral and robust variants:

\begin{itemize}
	\item \textbf{L1 and MSE (time-domain).} Basic waveform losses: mean absolute error (\texttt{l1\_loss}) and mean squared error (\texttt{mse\_loss}). These are stable and commonly used as a training baseline, particularly for models that predict waveforms or time-domain representations.

	\item \textbf{Multi-Resolution STFT loss} (\texttt{multistft\_loss}) \cite{auraloss}. A multi-scale spectral loss based on \texttt{auraloss.freq.MultiResolutionSTFTLoss}: the signal is compared in the frequency domain simultaneously across multiple STFT configurations (different window sizes and hop lengths). This improves perceptual quality and the recovery of high-frequency details, particularly for percussion and transients.

	\item \textbf{Robust quantile-masked MSE} (\texttt{masked\_loss}). A robust variant of MSE in which element-wise errors are computed and then the largest errors are discarded according to a quantile $q$ (trimmed/quantile loss). This reduces the influence of outliers and difficult segments. The quantile threshold is configurable via \texttt{training.q} in the YAML config.

	\item \textbf{Spectral RMSE} (\texttt{spec\_rmse\_loss}). The loss is computed after applying the STFT to both the prediction and the target: the complex spectrum is represented as a (Re, Im) pair, MSE is computed, and the square root is taken. This forces the model to match the spectrum in complex form rather than only in magnitude.

	\item \textbf{Robust spectral loss} (\texttt{spec\_masked\_loss}). Analogous to \texttt{masked\_loss} but in the complex STFT domain: spectral element-wise MSE is computed, after which the largest errors are discarded by quantile. This combines spectral comparison with robustness to outliers and anomalous segments.

	\item \textbf{LogWMSE} (\texttt{log\_wmse\_loss}). A loss based on the LogWMSE metric (implemented via \texttt{torch\_log\_wmse}). Unlike standard L1/MSE, it incorporates not only the prediction and target but also the input mix $x$, and operates in a perceptually weighted log-space. It also mitigates issues on silent portions of the track.
\end{itemize}

The final loss function in MSST is a weighted sum of selected components,
allowing the balance between waveform accuracy and spectral quality to be tuned 
for a specific architecture and dataset.

For models of the RoFormer and Conformer families (specifically the \texttt{bs} and \texttt{mel\_band} variants), MSST employs a specialised loss scheme. After the complex mask is predicted in the STFT domain, signals are reconstructed via the inverse STFT, and an L1 loss is computed between the reconstructed audio and the target stems. A multi-resolution STFT term is added: the sum of L1 divergences between STFT($\hat{y}$) and STFT($y$) across multiple window configurations, scaled by a \texttt{multi\_stft\_resolution\_loss\_weight} coefficient. The \texttt{active\_stem\_ids} mechanism allows training on a subset of stems rather than all simultaneously, which is particularly useful for sparse or multi-class datasets.

\subsection{Evaluation Metrics}

We support a range of metrics that are used to evaluate MSS tasks. 
Various metrics capture different aspects of separation quality, 
such as \emph{interference} (energy from non-target sources leaking into the estimate), 
\emph{artifacts}, \emph{distortion}, and \emph{perceptual quality}. 
No single metric covers all of these axes well, which is why our platform supportes a wide range of 
such complementary metrics, as described below. Taken together, these metrics cover complementary aspects of separation quality.

\begin{itemize}
	\item \textbf{SDR (Signal-to-Distortion Ratio)} is defined as the ratio of target signal power to total error 
	(leakage from other sources, artefacts, and noise) measured in decibels; 
	higher SDR indicates cleaner extraction.

\begin{equation}\label{eq:eq1}
SDR_{stem} = 10 \cdot \log_{10} \left( \frac{\sum\limits_{n=1}^{N} s_{stem,n}^2}{\sum\limits_{n=1}^{N} e_{stem,n}^2} \right)
\end{equation}
where $s_{stem,n}$ is the waveform of the ground truth, and $e_{stem,n}$ denotes the waveform of the estimate. The higher the SDR score, the better the output of the system.
 
SDR measures how well the desired audio sources have been separated 
from the mixture while minimizing the distortion caused by residual 
interference. To rank the entire system, the average SDR across all stems is used for each record.

The SDR metric is widely used to compare different sound demixing models in challenges~\cite{mdx23,MDX21}. 
Hovewer, it has several notable limitations: SDR gives significantly more weight to low frequencies than to high frequencies. 
It does not necessarily correlate well with human perception of 
separation quality, and can be sensitive to scaling differences between the reference and estimated signals. 

\item \textbf{SI-SDR (Scale-Invariant SDR)}~\cite{sisdr} is a variant of SDR that is invariant to overall scale (loudness),
eliminating trivial normalisation differences. It is also suitable as a differentiable training loss.

\item \textbf{L1Freq} \cite{l1freq} is a spectral metric comparing the magnitude spectrograms of the reference
and estimated audio signals using the short-time Fourier transform (STFT)
and calculating the L1 loss between them. The result is scaled to the range of 
[0, 100] (higher is better). It is more sensitive to errors distributed across the 
frequency axis than time-domain SDR. L1Freq is useful for monitoring spectral 
differences and for measuring spectral distortion even when waveform metrics appear acceptable.

\item \textbf{AuraSTFT / AuraMRSTFT} \cite{auraloss} metrics are based on 
\texttt{auraloss}. AuraSTFT metric~\cite{auraloss} evaluates the spectral difference between the 
reference and estimated audio signals using STFT loss (with log-magnitude and spectral centroid). 
The AuraSTFT metric computes the STFT loss in both logarithmic and linear magnitudes 
and is commonly used to assess the quality of audio separation tasks. 
By combining linear- and log-magnitude terms, AuraSTFT is best suited to capture 
spectral artifacts and distortion across a wide dynamic range.

AuraMRSTFT metric~\cite{auraloss} is similar to the AuraSTFT metric but 
evaluates the spectral difference between the reference and estimated audio 
signals using Multi-Resolution STFT loss instead with multiple FFT and window 
configurations (mel scale, perceptual weighting). 
Both are normalised to the range of 
[0, 100].

\item \textbf{LogWMSE (Logarithmic Weighted Mean Squared Error)} is a metric that 
evaluates the quality 
of the estimated signal compared to the reference signal in the context of audio 
source separation. The result is given on a logarithmic scale, which helps in evaluating 
signals with large amplitude differences. We implenented the LogWMSE metric based on \texttt{torch\_log\_wmse}.

LogWMSE is designed to address several shortcomings of common audio metrics, 
most importantly the lack of support for digital silence targets. It uses not only the reference and estimate but also the mixture (input mix), 
enabling quality evaluation in the context of the blended signal rather than 
purely against the isolated reference.

\item \textbf{Fullness and Bleedless:} MSST implements specialised indicators 
reflecting the practical 
usability of a separation. \textit{Fullness} measures how complete the extracted 
stem sounds: whether it contains dropped fragments or overly quiet segments. 
\textit{Bleedless} evaluates the absence of leakage from other sources: 
a high bleedless score indicates that, for example, the vocal track contains 
minimal extraneous instruments. During validation, MSST computes SDR together with 
bleedless and fullness for each stem by default.
\end{itemize}

\subsection{Optimisers}

Training MSS models typically employs adaptive gradient descent methods, as audio
losses (particularly spectral ones) can produce multi-scale gradients and require
stable optimisation. \textbf{Adam/AdamW} \cite{kingma2015adam, loshchilov2019adamw} is the standard starting point; AdamW
with decoupled weight decay generally provides more stable regularisation.
In MSST, the optimiser is configurable and selected via \texttt{config.training.optimizer}.
Available options include \textbf{Adam} \cite{kingma2015adam}, \textbf{AdamW} \cite{loshchilov2019adamw}, \textbf{RAdam} \cite{liu2020radam},
\textbf{RMSprop}, \textbf{Prodigy} \cite{mishchenko2024prodigy} (via \texttt{prodigyopt}), and
\textbf{AdamW8bit} \cite{dettmers2022eightbit} (via \texttt{bitsandbytes}) for memory-efficient training.
\textbf{Muon} \cite{jordan2024muon} and \textbf{AdaGO} are additionally supported:
parameters are automatically split into two groups -- 
tensors with $\mathrm{ndim} \geq 2$ (weight matrices and convolutions) 
are optimised by the Muon branch, while parameters with $\mathrm{ndim} < 2$ (biases and normalisation parameters) are handled by an AdamW-like branch.

A practical recommendation is to start with AdamW and move to specialised variants 
(Muon/AdaGO, Prodigy, 8-bit AdamW) only when faster convergence or reduced VRAM 
consumption is required. When fine-tuning from a pretrained checkpoint, Adam 
with a learning rate on the order of $\sim 9 \times 10^{-6}$ is often effective, as it adapts the model without disrupting learned representations.

\subsection{Training Configuration and Stability}

MSST uses YAML configuration files for all model and training hyperparameters (templates are provided in the \texttt{configs/} folder), ensuring experiment reproducibility and ease of modification. Key considerations include:

\begin{itemize}
	\item \textbf{Batch normalisation.} 
	The framework uses mean-variance normalisation: the mean and standard deviation of the 
	input $x$ are computed, and both $x$ and the target $y$ are normalised using the same statistics. 
	This aligns input and target scales and typically improves gradient stability.

	\item \textbf{Batch size and gradient accumulation.} \texttt{training.batch\_size} is set in 
	the config and is directly constrained by available VRAM. When a model does not fit, 
	the batch size is reduced and gradient accumulation is used to preserve the effective batch 
	size without memory overflow.

	\item \textbf{Learning rate and scheduler.} Two options are supported: \textbf{ReduceLROnPlateau} 
	(reduces LR when metric progress stalls; parameterised by \texttt{training.patience} 
	and \texttt{training.reduce\_factor}) and a \textbf{linear scheduler with warmup} 
	(parameterised by \texttt{training.num\_warmup\_steps} and total step count). 
	Learning rate selection is critical for stability: 
	excessively large values cause divergence or NaN losses, 
	while excessively small values lead to slow convergence.

	\item \textbf{Chunk size.} The training segment length is 
	set as \texttt{config.audio.chunk\_size} and directly affects both memory 
	consumption and the temporal context available to the model. 
	If training exceeds VRAM limits, reducing \texttt{audio.chunk\_size} is the first adjustment 
	to make. Typical values: for SCNet -- $485,100$ samples ($\approx$11~s), for BS-RoFormer -- $352,800$-$617,400$ 
	samples ($\approx$8-14~s duration), MDX23C -- $\approx$6~s.

	\item \textbf{Loss stability.} Training can be noisy in early epochs; 
	gradient spikes are typically addressed by reducing the learning rate, configuring the 
	scheduler appropriately, and applying gradient clipping. If the loss diverges to NaN, a more 
	conservative learning rate and/or a change in loss component composition usually resolves the issue.
\end{itemize}

MSST supports both single-GPU training (\texttt{train.py}) and multi-GPU training. Multiple device IDs can be specified via \texttt{--device\_ids} (e.g., \texttt{--device\_ids 0 1}) for \textbf{DataParallel} mode. A more efficient option is distributed training with \textbf{DDP} (\texttt{train\_ddp.py}), in which separate processes are launched per GPU with synchronised gradients. Training via \textbf{HuggingFace Accelerate} (\texttt{train\_accelerate.py}) is also available; in practice it works reliably in Linux environments.

\section{Inference and Advanced Techniques}

Once a model is trained, applying it efficiently to real audio recordings, 
which are often much longer than training segments, becomes the next challenge. 
MSST provides several tools for efficient inference and for improving output quality.

\subsection{Sliding Window and Cross-Fading}

During inference, a long track is divided into overlapping segments of fixed length. 
The window size is set by the parameter \texttt{config.inference.chunk\_size} 
and the degree of overlap is controlled by the parameter \texttt{config.inference.num\_overlap}: 
larger values reduce the step between windows and increase overlap. 
To eliminate discontinuities at segment boundaries, each segment is multiplied by a window 
with smooth edges (fade-in/fade-out), the results are summed, and then normalised by the 
accumulated weight. Reflect padding is applied at the track edges to prevent quality degradation 
in the first and last segments.

\subsection{Test-Time Augmentation}

Test-time augmentation (TTA) in MSST averages model predictions over several simple transformations of the input mix. 
The current implementation uses two augmentations: stereo channel swap and polarity inversion 
(multiplication of the signal by $-1$). For each transformed input, separation is performed; 
the results are then mapped back to the original orientation and summed with the original prediction. 
The final result is obtained by averaging over the three runs (original plus two TTA variants), 
improving output robustness through transformation invariance at the cost of proportionally 
increased inference time. Effect of TTA on source separation quality is shown 
in Table~\ref{tab:tta}. 
The results indicate that TTA provides a small but consistent improvement in 
SDR across different models and stems, demonstrating its utility for enhancing separation performance.

\begin{table}[htbp]
  \centering
	\caption{Effect of TTA on source separation quality.
$\Delta$SDR is the absolute gain from enabling TTA.}
\label{tab:tta}
\begin{tabular}{l l r r r r c}
\toprule
\textbf{Model (stem)} & \textbf{Mode} & \textbf{SDR} & \textbf{Bleedless} & \textbf{Fullness} & \textbf{L1\_freq} & \textbf{$\Delta$SDR} \\
\midrule
\multirow{2}{*}{BS Roformer (vocals)}
  & No TTA & 12.33 & 39.19 & 18.12 & 41.37 & \multirow{2}{*}{$+0.03$} \\
  & TTA    & 12.36 & 39.06 & 18.19 & 41.42 & \\
\midrule
\multirow{2}{*}{SCNet XL (drums)}
  & No TTA & 13.36 & 24.14 & 21.32 & 40.22 & \multirow{2}{*}{$+0.00$} \\
  & TTA    & 13.36 & 24.13 & 21.32 & 40.23 & \\
\midrule
\multirow{2}{*}{MelBand Roformer (vocals)}
  & No TTA & 11.30 & 39.00 & 15.95 & 38.98 & \multirow{2}{*}{$+0.06$} \\
  & TTA    & 11.36 & 38.81 & 16.01 & 39.05 & \\
\midrule
\multirow{2}{*}{BS Roformer (keys)}
  & No TTA & 8.93 & 33.81 & 15.29 & 28.64 & \multirow{2}{*}{$+0.03$} \\
  & TTA    & 8.96 & 29.86 & 16.78 & 28.71 & \\
\midrule
\multirow{2}{*}{MDX23C (guitar)}
  & No TTA & 6.34 & 22.64 & 15.81 & 35.21 & \multirow{2}{*}{$+0.01$} \\
  & TTA    & 6.35 & 22.70 & 15.82 & 35.21 & \\
\bottomrule
\end{tabular}
\end{table}

\subsection{LoRA Adaptation}

LoRA (Low-Rank Adaptation) \cite{hu2021lora} enables fine-tuning of large models by keeping the base weights frozen and training only added low-rank matrices. This reduces the number of trainable parameters and memory requirements, enabling rapid adaptation of a trained MSS model to a new domain or dataset. MSST supports LoRA at the training pipeline level: it can be enabled via a flag at training launch and configured through the YAML config. The resulting LoRA weights are stored separately from the main model and can be applied on top of the base checkpoint during validation and inference, facilitating lightweight adaptations without full retraining.

\subsection{Flexible Checkpoint Loading}

MSST provides a mechanism for partial weight loading even when the model architecture has changed. Specifying \texttt{--start\_check\_point} with an existing weight file causes the code to load matching layers and ignore mismatched parameters. For example, Band-Split RoFormer training can be initialised from Mel-Band RoFormer weights: shared components (Transformer blocks) are loaded from the checkpoint, while architecture-specific parts are trained from scratch. This transfer accelerates convergence and improves training stability. A checkpoint file (\texttt{.ckpt}) stores the model \texttt{state\_dict}, optimiser state, current epoch, best achieved metrics, and auxiliary information. MSST automatically saves checkpoints per epoch and upon metric improvement.

\subsection{ONNX and TensorRT Export}

To accelerate inference and facilitate integration into third-party applications, the repository provides scripts for exporting MSST checkpoints to \textbf{ONNX} and \textbf{TensorRT} formats \cite{msst_onnx}. This conversion enables running models without a PyTorch dependency and achieves substantial speed gains on both CPU and GPU with TensorRT support.

\subsection{GUI and Practical Deployment}

MSST includes a graphical user interface that simplifies model usage for users without command-line experience. Through the GUI, a user can select a pretrained model, specify an input audio file, and launch separation with a few clicks, receiving individual stems as output. This mode is useful for quality assessment, model comparison, and demonstration to audio engineers or musicians. The GUI serves as a wrapper around the inference pipeline, enabling model evaluation and deployment without writing code.

\section{Model Ensembling}

Separation quality can be improved by combining the outputs of multiple models. 
MSST supports output ensembling (via the \texttt{ensemble.py} script) and 
integrates this capability into the GUI. In practice, ensembling consistently 
yields gains in the SDR metric, typically ranging from +0.01 to +0.2~dB when 
combining several top-performing models, at the cost of proportionally increased 
processing time.

Ensembling makes it possible to combine the output audio signals of several models
both in the time domain (sample-wise computation of the 1D signal) and in the
frequency domain (pixel-wise processing of spectrograms based on the short-time
Fourier transform, STFT). The aggregating functions used are the mean, the median,
as well as the minimum and the maximum, which allows the ensemble parameters to be
tuned flexibly. The choice of a particular function determines the properties of the
resulting signal: the spectral minimum, for instance, reduces the influence of overly
aggressive models and yields a more conservative result, whereas the spectral maximum
behaves most aggressively. The key prerequisite for the successful application of this
approach is a comparable baseline quality of the models being combined. Empirical
studies have shown that the optimal signal-to-distortion ratio (SDR) values are
consistently achieved with weighted averaging of the signals directly in the time domain.

\begin{table}[htbp]
  \centering
  \caption{Composition of the models ensembles and SDR values.}
  \label{tab:ensembles}
  \begin{tabular}{llccc}
    \toprule
    \textbf{Source type} & \textbf{Models in the ensemble} & \textbf{Weight} &
    \textbf{Model SDR} & \textbf{Ensemble SDR} \\
    \midrule
    \multirow{3}{*}{Vocals}
      & BSRoformer vocals      & 40 & 11.24 & \multirow{3}{*}{11.61} \\
      & MelBandRoformer vocals & 40 & 11.23 & \\
      & SCNet XL               & 25 & 10.96 & \\
    \midrule
    \multirow{2}{*}{Wind}
      & SCNet Large            & 10 & 6.76  & \multirow{2}{*}{7.22} \\
      & MelBand Roformer       & 10 & 6.52  & \\
    \midrule
    \multirow{2}{*}{Crowd}
      & Crowd model MDX23C     & 10 & 6.06  & \multirow{2}{*}{6.27} \\
      & MelBand Roformer       & 10 & 6.07  & \\
    \midrule
    \multirow{2}{*}{Bass}
      & SCNet XL               & 11 & 13.81 & \multirow{2}{*}{14.87} \\
      & BS Roformer            & 24 & 14.62 & \\
    \bottomrule
  \end{tabular}
\end{table}

\section{Community-Developed Models}
\label{sec:community-models}

The availability of a unified training and evaluation framework~\cite{msst_github} 
has enabled the community to train and release a substantial number of 
open models beyond the canonical
\emph{bass}, \emph{drums}, and \emph{vocals} stems. They include 
specialised models for specific separation tasks 
such as aspiration extraction, phantom-centre extraction, dereverberation,
denoising, and others. This ecosystem of community-contributed models demonstrates the 
practical utility of the MSST framework as a platform for both research and 
applied audio 
engineering. The most widely adopted contributions are summarised
below.

\begin{enumerate}
    \item \textbf{Vocal/instrumental separation.} Kimberley Jensen released
    a vocal-instrumental model~\cite{jensen2023melband} built on the
    Mel-Band RoFormer architecture \cite{wang2023mel}.

    \item \textbf{Crowd extraction.} The developer of Ultimate Vocal
    Remover (UVR)~\cite{uvr}, working under the alias
    \texttt{aufr33}, together with \texttt{viperx}, published model for
    the crowd-noise extraction task~\cite{aufr33crowd}. These model weights were
    subsequently converted to the OpenVINO runtime for optimised CPU
    inference~\cite{intel2024crowdopenvino}.

    \item \textbf{Dereverberation.} Several model checkpoints targeting the
    removal of reverberation were released by the Discord community \texttt{Audio Separation}.

    \item \textbf{Denoising.} Dedicated model weights for broadband noise
    removal~\cite{msst2024denoise}.

    \item \textbf{Aspiration extraction.} A model isolating the
    \emph{aspiration} component of a vocal track (the breathy,
    whisper-like residue of singing) was contributed by
    SUC-DriverOld~\cite{sucial2024aspiration}. The model separates three
    perceptually distinct phenomena:
    \begin{enumerate}
        \item audible breathing;
        \item the hiss and buzz of fricative consonants (e.g.\ \textipa{/s/}
        and \textipa{/f/});
        \item plosives, i.e.\ the unvoiced burst produced when articulating
        stop consonants such as \textipa{/p/}, \textipa{/t/}, and
        \textipa{/k/}.
    \end{enumerate}

    \item \textbf{Phantom-centre extraction.} Weights for isolating the
    phantom centre were released in a later revision of the model
    collection~\cite{msst2024phantom}. The phantom centre is a
    psychoacoustic illusion: it arises when the left and right channels
    reproduce an identical monophonic signal at equal amplitude and in
    phase. The auditory system integrates the two coherent inputs and
    localises a virtual source directly ahead of the listener, despite no
    physical transducer being present at that position.

    \item \textbf{Drum-kit decomposition.} A number of checkpoints
    decomposing the drum stem into its constituent instruments~\cite{politrees2024drums} were trained
    for both the HTDemucs~\cite{rouard2023htdemucs} and
    MDX23C~\cite{MDX23C} architectures.
\end{enumerate}

A curated index of the best-performing community models checkpoints is maintained
in the project repository~\cite{msst2024pretrained}.

\section{Conclusion}

For MSS tasks, practical training and evaluation pipeline configuration, 
such as model architecture, data preparation and 
augmentation, loss function design, 
evaluation metrics, training stability, and optimised inference on 
long tracks, play an important role alongside architectural 
innovations in achieving best possible results.

MSST addresses this full cycle by providing a reproducible, 
configuration-driven infrastructure for training and evaluating demixing 
models within a single unified framework. Its primary strengths are 
modularity and 
applied orientation. We provide support for multiple modern model families, 
seven dataset types and a 
rich augmentation pipeline, flexible composition of weighted loss functions 
and a range of evaluation metrics, 
single-GPU and multi-GPU training, as well as 
built-in techniques that deliver reliable quality improvements without 
model modifications (sliding-window inference with cross-fading, TTA, 
model ensembling, and LoRA adaptation).

As a result, MSST serves both as a research platform for systematic ablation 
studies and as a practical tool that enables rapid progression from idea 
to verifiable result without the overhead of building training 
infrastructure from scratch. The growing ecosystem of community-contributed 
models further validates the framework's utility for the broader audio 
engineering community.

\section*{Acknowledgements}

This research was supported in part through computational resources of HPC facilities at HSE University \cite{kostenetskiy2026hpc}.

\bibliographystyle{unsrtnat}
\bibliography{references}

\end{document}